\documentclass[12pt,preprint]{aastex}
\newcommand{\Msun}{~M_\odot}

\newcommand{\gcm}{\rm ~g~cm^{-3}}
\newcommand{\kms}{\rm ~km~s^{-1}}
\newcommand{\ergs}{\rm ~erg~s^{-1}}

\newcommand{\ml}{~\Msun ~\rm yr^{-1}}

\begin{document}

\title{CIRCUMSTELLAR EMISSION FROM TYPE Ib AND Ic SUPERNOVAE}
\author{
Roger A. Chevalier\altaffilmark{1} and
Claes Fransson\altaffilmark{2}
}
\altaffiltext{1}{Department of Astronomy, University of Virginia, P.O. Box 400325, 
Charlottesville, VA 22904-4325; rac5x@virginia.edu}
\altaffiltext{2}{Department of Astronomy, Stockholm University, AlbaNova, 
SE--106~91 Stockholm, Sweden}

\begin{abstract}
The presumed Wolf-Rayet star progenitors of Type Ib/c supernovae
have fast, low density winds and the shock waves generated
by the supernova interaction with the wind are not expected
to be radiative at typical times of observation.
The injected energy spectrum of radio emitting
electrons typically has an observed index
$p=3$, which is suggestive of acceleration in cosmic ray dominated shocks.
The early, absorbed part of the radio light curves can be attributed
to synchrotron self-absorption, which leads to constraints on the
magnetic field in the emitting region and on the circumstellar density.
The range of circumstellar densities inferred from the radio emission
is somewhat broader than that for Galactic Wolf-Rayet stars, if
similar efficiencies of synchrotron emission are assumed in the extragalactic
supernovae.
For the observed and expected ranges of circumstellar densities to roughly
overlap, a high efficiency of magnetic field production in the shocked region
is required ($\epsilon_B\approx 0.1$).
For the expected densities around a Wolf-Rayet star, a nonthermal mechanism
is generally required to explain the observed X-ray luminosities of
Type Ib/c supernovae.
Although the inverse Compton mechanism can explain the observed X-ray
emission from SN 2002ap if the wind parameters are taken from the
radio model, the mechanism is not promising for other supernovae unless
the postshock magnetic energy density is much smaller than the electron
energy density.
In some cases another mechanism is definitely needed and
we suggest that it is X-ray synchrotron emission in a case where the shock
wave is cosmic ray dominated so that the electron energy spectrum flattens
at high energy.
More comprehensive X-ray observations of a Type Ib/c supernova are
needed to determine whether this suggestion is correct.

\end{abstract}

\keywords{stars: circumstellar matter --- stars: mass loss --- supernovae}

\section{INTRODUCTION}

SNe Ib/c (Type Ib and Ic supernovae) are hydrogen free, or nearly 
hydrogen free,
and are thought to have stripped massive star core,
i.e. Wolf-Rayet star, progenitors.
The observed rate of SNe Ib/c indicates that they make up about
1/4 of massive star supernovae \citep[e.g.,][]{DF99}
and thus are an important mode
of massive star death.
Although single very massive stars may end their lives as Wolf-Rayet stars,
they are probably not in sufficient numbers to account for
all of the SNe Ib/c \citep{WL99};
binary progenitors are likely to be important contributors to the rate.

Considerable interest in SNe Ib/c has been generated by their
relation to GRBs (gamma-ray bursts).
The spatial and temporal correlation of the Type Ic
SN 1998bw with the burst GRB 980425 led to the probability
of a chance superposition of only $10^{-4}$ \citep{Gal98},
and the extraordinary energy displayed by the optical \citep{Iwa98}
and radio \citep{Kul98} emission from the supernova
made the identification even more secure.
The case was clinched by the finding of Type Ic supernovae similar
to SN 1998bw in the cosmological GRBs 030329 and 031203
\citep{Sta03,Mal04}.
The broad line character of the optical spectrum of SN 1998bw,
with velocities up to $60,000\kms$, led
to the view that SNe Ib/c with such broad lines were
likely to be related to GRBs,  
including SN 2002ap \citep{Maz02}.
However, SN 2002ap
was a relatively weak radio \citep{BKC02} and X-ray source.
On the basis of the radio emission, \cite{BKC02} argued
that the energy in high velocity ejecta was low, so that
a connection to GRBs was unlikely for this object.

Although it is not known which SNe Ib/c have a GRB connection,
it is clear at least some of them do and this
led to the prediction that
late radio observations of SNe Ib/c could yield the detection of
an off-axis GRB jet that slowed down and radiated more isotropically
\citep{Pac01}.
Searches for such emission were undertaken \citep{Sto03,Ber03,Sod06a},
leading to the detection of luminous radio and X-ray emission
from SN 2001em \citep{Sto04},
SN 2003L \citep{Sod05}, and SN2003bg \citep{Sod06b}.
In the case of SN 2003L, broad-lined optical
emission is not present \citep{Mat03}, and in no case is there
direct evidence for a GRB connection.
SNe Ib/c  were generally not detected
at radio wavelengths, showing
that such luminous supernovae are rare \citep{Sod06a}.

Although there have been discussions of the radio and X-ray
emission from individual SNeIb/c, there has been little discussion
of their overall properties and how they relate to the mass
loss properties of the progenitor star.
We consider these properties and discuss the mechanisms for X-ray
emission from the supernovae.
In \S~2, the observed properties of SNe Ib/c are briefly reviewed.
The hydrodynamics of the interaction and the resulting velocities are
estimated in \S~3.
The expected relativistic electron spectrum, allowing for loss
processes, is treated in \S~4 and the resulting synchrotron radio
emission in \S~5.
The possible mechanisms for the X-ray emission and their application
to the observations are discussed in \S~6.
A discussion of the results is in \S~7.

\section{OBSERVED PROPERTIES}

Clear signs of circumstellar interaction in SNe Ib/c are not
present at optical wavelengths, but are seen at radio and
X-ray wavelengths.
Radio light curves have been
observed for SN 1983N \citep{Wei86},
SN 1984L \citep{PSW86},
SN 1990B \citep{Van93},
SN 1994I \citep{Sto05}, SN 1998bw \citep{Kul98},
SN 2001ig \citep{Ryd04},
SN 2002ap \citep{BKC02},
SN 2003L \citep{Sod05},
and SN 2003bg \citep{Sod06b}.
SN 2001ig was observed to have H lines in its spectrum at one point,
but some of its properties point to a Wolf-Rayet star progenitor
\citep{Ryd04,Sod06b}, so we include it here.

The light curves for the radio supernovae show an early increase
followed by a decline.
The peak spectral luminosity and time of peak for various supernovae
are shown in Fig. 2 of \cite{CFN06}.
If SSA (synchrotron self-absorption) is the dominant absorption process,
the position of the supernovae in this plot gives an indication of the
velocity of the radio emitting shell \citep{Che98}.
SNe Ib/c generally separate themselves from other
supernovae in this plot by having high velocities.
The Type Ic SN 1998bw, associated with GRB 980425, stands out even more
by requiring relativistic motion.
Other Type Ic supernovae associated with GRBs, e.g., SN 2003dh with
GRB 030329 and SN 2003lw with GRB 031203, had radio properties
indicating highly relativistic motion.
In these last two cases, the radio emission is generally 
believed to be associated
with power from the same central engine that caused the GRB.
The case of SN 1998bw has been more controversial.  \cite{LC99} suggest
an association with the GRB power source,
but others have argued for a supernova origin for the radio
and X-ray emission \citep{Wax04}.
Other than SN 1998bw, we do not treat the SNe Ic associated with GRBs here.

One quantity of interest that can be found from radio observations
in the optically thin regime is the energy spectral index of the
synchrotron emitting electrons.
The values of $\alpha$, where $F_{\nu}\propto \nu^{\alpha}$, are
given in Table 1.
These values cover a range of times, but go to as late as several
hundred days in some cases.
The corresponding energy spectral index is $p=-2\alpha +1\approx 3$
for these supernovae.
The spectral indices are generally steeper for the SNe Ib/c than
for other types of supernovae.
Table 1 also gives $\beta$, where $F_{\nu}\propto t^{\beta}$, in
the optically thin regime.
The uncertainties in these values are $\sim 0.1-0.2$ and vary
among the different objects.
There are only a few data points for SN 1984L, so those results
are especially uncertain.
SN 2001ig and SN 2003bg both showed jumps in their light curves;
the numbers used here do not include those parts of the light curves.

SNe Ib/c generally have no more than a few X-ray observations, so
there is no complete light curve information.
Table 2 lists the current observations, including the unabsorbed
luminosities, of SNe Ib/c;
all the detections are  of Type Ic objects, which are the ones without
He lines.
The distances in Table 2 and 3 are based on $H_0=70$ km s$^{-1}$ Mpc$^{-1}$.
For SN 2003L and SN 2003bg, the distances are based on the recession
velocity of the host galaxy and this value of $H_0$;
in all the other cases, the distances of \cite{TF88} have been
scaled to $H_0=70$ from their assumed value of 75.
The resulting distance to SN 2002ap, 10.4 Mpc, is at the high end of
the range of distances typically chosen for this supernova, but
we are especially interested in having a consistent set of relative distances.
SN 1994I in M51 was detected by {\it Chandra} at an age of 
$\sim 7$ years \citep{Imm02};
apparently it was detected at a later age in X-rays than in any other
wavelength band.

\section{HYDRODYNAMICS}

The hydrodynamic interaction between a supernova and its surroundings
depends on the density structure of the freely expanding supernova
ejecta and the structure of the circumstellar medium.
If SNe Ib/c have Wolf-Rayet progenitors,
typical wind properties are
$\dot M \approx 10^{-5}\ml$ and  $v_w\approx 1000\kms$.
We define a constant to specify the wind density $\rho=Ar^{-2}=
\dot M/4\pi r^2v_w$, with a reference value
$A_*=A/(5\times 10^{11}~{\rm g~cm^{-1}})$
corresponding to the typical wind properties.
Because of the high wind velocity, the supernova shock samples a
region of recent mass loss, so the assumption of steady mass
loss and a $r^{-2}$ density profile is reasonable.
For the mass loss rates and wind velocities determined by
\cite{NL00} for Galactic Wolf-Rayet stars, $A_*$
ranges from $0.07-7.4$.
The low value is for a WO star with $v_w=5,500\kms$.
If the supernova shock moves out at $30,000\kms$ and the
wind velocity is $1000\kms$, then the age of the wind that
is sampled by the supernova is $30t$, where $t$ is the age
of the supernova.
Since $t$ is typically $<$ a few years for observed SNe Ib/c,
the radio emission is sensitive to changes in the mass loss
properties on only a short timescale before the explosion
($\la 10^2$ yr).
If a slow wind were present from an earlier evolutionary phase,
a longer timescale for change would be possible because of the
slower wind velocity; this
may have occurred in the Type Ib/c SN 2001em \citep{CC06}.
We do not discuss that situation any further here.

A number of studies of SNe Ib/c have quoted mass loss rates assuming
a wind velocity of $10\kms$ 
\citep[e.g.,][]{Van93,Imm02,Ryd04}.
A rationale for a low velocity is that the wind is from a late type
companion with a slow wind.
However, the ram pressure of a Wolf-Rayet star wind is likely to dominate
that of any companion star.
\cite{Van93} address this by suggesting that SNe Ib/c are the
thermonuclear explosions of cores.
The further study of SNe Ib/c has supported the Wolf-Rayet hypothesis for
their progenitors and we assume a fast wind here.
When they occur in binary systems, the Wolf-Rayet wind can sweep back
the companion star wind.
The spiral structure that then results from the binary motion
has been observed in some Wolf-Rayet systems \citep{TMD99}.

The supernova structure depends on explosion models.
\cite{MM99} discussed the
density profiles resulting from the explosions of stars
with radiative envelopes, as expected for SNe Ib/c progenitors.
Radiative losses at the time of shock breakout
set the upper limit to the velocity of the material that is
shock accelerated at the outer edge of the star.
The maximum velocity can be expressed as
\begin{equation}
v_{fmax}=115,000 ~E_{51}^{0.58}\left(M\over 10\Msun\right)^{-0.42}
\left(R_*\over 1~R_\odot\right)^{-0.32}\kms,
\label{vmax}
\end{equation}
where $E$ is the explosion energy in units of $10^{51}$ ergs, $M$ is the
ejecta mass, $R_*$ is the progenitor radius ($1~R_\odot$ is typical for a
Wolf-Rayet star), and
typical values have been used for the opacity and the progenitor
density structure.

The resulting density profile, approximately a steep outer power
law and a shallow inner power law, can be scaled with the
ejecta mass, $M$, and energy $E$.
Models of the optical light curves and spectra
of particular supernovae  lead to estimates
of $M$ and $E$.
The outer density structure can be written as \citep{BKC02}
\begin{equation}
\rho_{sn}=3\times 10^{96}E_{51}^{3.59}\left(M\over 10\Msun\right)^{-2.59}
t^{-3}v^{-10.18},
\end{equation}
where $v=r/t$ for the freely expanding gas and cgs units are used
for the coefficient.

The interaction of the steep outer power law structure with  the
surrounding   wind can be described by a self-similar solution
for a $\gamma=5/3$ fluid \citep{Che82a}.
The solution should apply at early times and can be used to find
the radius of the contact discontinuity between the ejecta and the
circumstellar gas, $R_c$,
\begin{equation}
{R_c\over t}=3.0\times 10^4 E_{51}^{0.43}\left(M\over 10\Msun\right)^{-0.32}
A_*^{-0.12}t_{10}^{-0.12}\kms,
\label{rsh}
\end{equation}
where $t_{10}$ is the age in units of 10 days.
The forward shock is at $1.24R_c$ and the reverse shock at $0.984R_c$.
The density at the reverse shock is 29 times that at the forward shock
and the mass of shocked ejecta is 2.0 times that of swept up
circumstellar medium.
In this solution, the fraction of the volume within the outer shock wave
occupied by shocked gas
is $f=0.50$, which is made up of
shocked circumstellar gas ($f=0.47$) and shocked ejecta ($f=0.03$).

Combining equations (\ref{vmax}) and (\ref{rsh}) gives the approximate time at
which interaction with the power law profile begins
\begin{equation}
t_{m}=0.0014 E_{51}^{-1.17}\left(M\over 10\Msun\right)^{0.83}A_*^{-1}
\left(R_*\over 1~R_\odot\right)^{2.67}~\rm day.
\end{equation}
This is earlier than the time at which radio supernovae are observed,
so the initial hydrodynamic evolution is expected to be in the self-similar
regime of interaction with a steep power law.
Deviations from self-similar behavior are expected when the reverse
shock front moves into the interior, flatter part of the supernova
density profile.
To estimate the importance of this effect, we use the harmonic
mean density distribution of \cite{MM99}.
Results for a particular density profile can be scaled to any appropriate
values of stellar mass, $M$,  explosion energy, $E$, and
circumstellar density, $A$, using the
scaling laws  that lengths $\sim M/A$ and time $\sim M^{3/2} E^{-1/2}
A^{-1}$, so that $v\propto (E/M)^{1/2}$, $\rho\propto (A^3/M^2)$, and
$p\propto A^3 E/M^3$.
We use results for the interaction in the thin shell approximation,
as described by \cite{Che05}.
These results show that at an age
\begin{equation}
t_e=19 E_{51}^{-0.5}\left(M\over 10\Msun\right)^{1.5}
A_*^{-1}~\rm yr,
\end{equation}
the value of $m=(dR/dt)t/R=0.866$, corresponding to $n=9.5$.
This value is close to the initial value, so that the assumption
of self-similar evolution over the time of observation of radio
supernovae (at most up to a few years old) is a reasonable approximation.

The density structure advocated here can be compared to that found by
other groups.
The model CO100 has been used by Mazzali and collaborators to
model energetic Type Ic supernovae 
\citep[see Fig. 8 of][ for the density structure]{MIN00}.
The high velocity ($>25,000\kms$) end of this structure can be
approximated by a steep power law ($\rho\propto r^{-9.0}$), in
reasonable agreement with the model used here.
However, \cite{MIN00} find that the spectroscopic modeling
of the broad-line Type Ic SN 1997ef is improved by having a relatively
flat density profile ($\rho\propto r^{-4}$) over the velocity
range $25,000-50,000\kms$.
If the reverse shock wave of the interaction region propagated into
such a shallow density gradient, the forward shock wave would begin
to evolve toward $R\propto t^{2/3}$ blast wave expansion.

Although the self-similar solution is expected to be adequate for
the typical case, there may be circumstances where it fails.
One is if there is a combination of high $E$, low $M$, and low $A_*$
so that the flow is relativistic; 
this requires $\beta\Gamma \approx 1$, where $\beta=v/c$ and
$\Gamma=(1-\beta^2)^{-1/2}$,
or $v\approx 210,000\kms$ in equation (\ref{rsh}).
\cite{TMM01} have discussed the supernova structure
for the case where the accelerating shock at the outer edge of the
star becomes relativistic.
However, equation (\ref{rsh}) shows that even for $E_{51}=10$
the shock velocity is not relativistic for radio and X-ray supernovae
at typical times of observation.
Another circumstance is recent change from a different stellar
evolutionary phase, as described above.

Deviations from the $\gamma=5/3$ energy-conserving
self-similar solution are expected if radiative cooling
of the gas is important.
For the hydrodynamic model described above, the velocity of gas moving
into the reverse shock is
\begin{equation}
V_{rs}=3.36\times 10^3 E_{51}^{0.43}\left(M\over 10\Msun\right)^{-0.32}
A_*^{-0.12}t_{10}^{-0.12}\kms,
\end{equation}
and the preshock density is
\begin{equation}
\rho_{rs}=1.44\times 10^{-18}E_{51}^{-0.87}\left(M\over 10\Msun\right)^{0.63}
A_*^{1.24}t_{10}^{-1.76} \gcm.
\end{equation}
Unless the ejecta are very heavy element
rich, the radiative cooling is dominated by
free-free emission (see \S~6.1) and the cooling time is larger than the age
of the supernova even at $t=1$ day.
The temperature gradually declines with time, but the density also
declines and the cooling time remains larger than the age.
The temperature at the outer shock is higher and density lower,
so the cooling times are even longer.
This justifies the use of the nonradiative self-similar solution.

Another effect that can change the structure is the efficient
acceleration of relativistic particles.
If relativistic particles are an important contributor to
the pressure, the adiabatic index approaches $4/3$ and
the compression ratio in the shock front is increased.
The result is a decrease in the width of the shocked
region \citep{Che83}.
If there are losses of relativistic particles from the
shocked region, the compression in the shocked region is
increased further \citep[e.g.,][]{ER91}.

\section{RELATIVISTIC PARTICLE SPECTRUM}
\label{sp}

Particle acceleration is most likely to occur at the shock fronts
resulting from the interaction.
There are reasons to believe that the forward shock dominates
the reverse shock in contributing to the relativistic particle component.
The forward shock has a higher velocity, postshock energy density,
and volume of shocked gas.
In addition, the preshock magnetic field at the reverse shock
is small because of the expansion of the supernova.
In the case of Tycho's supernova remnant, \cite{War05} have
argued on the basis of X-ray observations that the particle
acceleration is primarily at the forward shock and not
at the reverse shock.

Loss processes that can affect the particle distribution
are synchrotron losses and inverse Compton losses.
We assume that a fraction $\epsilon_B$ of $\rho_0 v_{s}^2$,
where $\rho_0$  is the preshock circumstellar density and
$v_s$ is the forward shock velocity,
is converted to magnetic energy density, $u_B$, so that
\begin{equation}
u_B=0.052 \epsilon_{B-1}A_* t_{10}^{-2}
~{\rm erg~cm^{-3}},
\label{ub}
\end{equation}
where $\epsilon_{B-1}$ is $\epsilon_{B}/0.1$ and $t_{10}$
is the age in units of 10 days.
The value of $B$ for the reference parameters is 1.1 G.
Assuming that electrons radiate at their critical synchrotron
frequency, the Lorentz factor of electrons radiating at
frequency $\nu$ is
\begin{equation}
\gamma=56\nu_{10}^{0.5}\epsilon_{B-1}^{-0.25}A_*^{-0.25}t_{10}^{0.5},
\label{gam}
\end{equation}
where $\nu_{10}=\nu/10^{10}{\rm~GHz}$.
The synchrotron cooling time can then be written as
\begin{equation}
t_{syn}=14\nu_{10}^{-0.5}\epsilon_{B-1}^{-0.75}A_*^{-0.75}t_{10}^{1.5}~{\rm day}
\label{ts}
\end{equation}
and the critical frequency at which the synchrotron cooling time
equals the age is
\begin{equation}
\nu_{c10}=1.8\times 10^2 \epsilon_{B-1}^{-1.5}A_*^{-1.5}t_{10}.
\end{equation}
It can be seen that synchrotron cooling is typically not important
when considering radio emission, but is important when considering
X-ray emission.
If the electron shock acceleration extends to a high energy, the
effect of synchrotron cooling is to steepen the energy spectrum
from $p$ to $p+1$.

As discussed by \cite{BF04}, inverse Compton losses
can be significant when considering the radio emission from a Type Ic
supernova.
The radiation field is dominated by that from the supernova photosphere,
which gives an energy density $u_{rad}=L/(4\pi R^2 c)$, where $L$ is
the supernova luminosity.
Consideration of the light curve shape for a SN Ib/c shows that the
Compton cooling is most important close to maximum light.
For events like SN 1994I and SN 2002ap, the luminosity peaked at
$\sim 2\times 10^{42}\ergs$, when the age was 10 days.
The Compton cooling time can be written as
\begin{equation}
t_C={25.2\over u_{rad}E}=16\nu_{10}^{-0.5}\epsilon_{B-1}^{0.25}
A_*^{0.01}t_{10}^{1.26}
E_{51}^{0.88}M_1^{-0.64}\left(L\over 2\times 10^{42}\ergs\right)^{-1}
~{\rm day},
\label{tc}
\end{equation}
where cgs units are used in the first expression.
This result shows that Compton cooling can have some effect near maximum
light, as was found by \cite{BF04} for SN 2002ap,
but that it is probably not important at later times.

  From equations (\ref{ts}) and (\ref{tc}), the ratio of synchrotron cooling time to inverse
Compton cooling time is
\begin{equation}
{t_{syn}\over t_{IC}}=0.9 \epsilon_{B-1}^{-1.0}A_*^{-0.76}t_{10}^{0.24}
E_{51}^{-0.88}M_1^{0.64}\left(L\over 2\times 10^{42}\ergs\right).
\label{stoi}
\end{equation}
After maximum light, the luminosity $L$ drops more rapidly than $t^{-0.24}$,
so synchrotron cooling is expected to be the dominant cooling
process.

Assuming that the
maximum energy to which electrons are accelerated, $E_{max}$, is limited
by synchrotron losses, we have \citep[e.g.,][]{Rey98}
\begin{equation}
E_{max}\propto B^{-1/2}v_{s}G^{1/2},
\label{emax}
\end{equation}
where $v_s$ is the forward shock velocity,
and $G$ is a factor to account for inverse Compton losses
and is $\propto u_B/u_{rad}$ when the inverse Compton losses dominate
the synchrotron losses.
The constant of proportionality in equation (\ref{emax}) depend on
details of the shock acceleration process and the obliquity of
the shock front.
Taking the synchrotron emission to be at the critical frequency,
the maximum frequency for the synchrotron emission is
\begin{equation}
\nu_{max}\propto BE^2\propto v_{s}^2 G.
\end{equation}
X-ray synchrotron emission has been observed from Galactic
supernova remnants like SN 1006 with $v_{s}\approx 3000 \kms$
\citep{Koy95}, so
the high velocities present in young supernovae should produce
emission beyond the X-ray regime if synchrotron losses dominate.
The above considerations show that synchrotron losses typically
do dominate, except near maximum optical light when $G\sim 0.1$
for the standard parameters.
Even in this case, the emission should extend to X-ray wavelengths.
The synchrotron cooling time is much less than the age at these
high frequencies, so the supernova age is not a limiting factor
in determining the maximum particle energy.

As noted in \S~2, radio observations of SNe Ib/c indicate
an energy spectral index $p\approx 3$ over a broad range of times.
The fact that synchrotron and inverse Compton losses are unlikely
to affect the spectrum implies that this is the injection spectrum
for the radiating particles.
Diffusive shock acceleration in the test particle limit is
well known to give the result $p=(r+2)/(r-1)$, where $r$ is
the shock compression.
A nonrelativistic shock with $r=4$ gives $p=2$.
A steeper spectrum is possible if there is feedback to the acceleration
process and the lower energy particles cross a subshock with a relatively
low degree of compression \citep[e.g.,][]{ER91,EBB00}.
Higher energy particles experience a higher degree of compression and
thus have a flatter spectrum; there is the possibility
of curvature in the spectrum.
The total compression can be $r>4$ (leading to $p<2$) because of the
relativistic contribution to the equation of state and the loss of
high energy particles from the shock region.

At the high shock velocities in SNe Ib/c,
many of the postshock electrons can be accelerated
to relativistic velocities.
If a fraction $\epsilon_{et}$ of the total postshock energy density
goes into electrons then (cf. Soderberg et al. 2005)
\begin{equation}
\epsilon_{et} \left({v\over c}\right)^2 {m_p c^2
\over r}\approx \bar\gamma m_e c^2,
\label{epe}
\end{equation}
where $\bar\gamma$ is the average Lorentz factor of the postshock
electrons.
The electron distribution behind fast shocks is not well understood
theoretically.
In the field of GRBs, it is commonly assumed that all the electrons
go into a power law spectrum with energy index $p$, above some
Lorentz factor $\gamma_{min}$.
If a similar assumption is made here, then
\begin{equation}
\bar\gamma=(p-1)\gamma_{min}/(p-2);
\label{gbar}
\end{equation}
in this case $\epsilon_e=\epsilon_{et}$,
where $\epsilon_e$ is the fraction of the postshock energy density
that goes into nonthermal relativistic electrons.
The requirement that $\gamma_{min}> 1$ gives
\begin{equation}
\epsilon_e > 0.16 \left(v_{s}\over 50,000\kms\right)^{-2}
\label{rel}
\end{equation}
for $p=3$ and we have assumed that low energy particles dominate
the electron energy.
The postshock energy is distributed among the electrons, magnetic
field, and protons/ions, so $\epsilon_e\la 0.33$ is likely.
If the constraint in equation (\ref{rel}) is not fulfilled, 
there is a significant nonrelativistic electron
component.

\section{RADIO EMISSION}

\label{sradio}

The basic physical mechanisms relevant to the radio emission from
SNe Ib/c have been discussed by \cite{Che98} and \cite{BF04}.
There is an early phase when synchrotron self-absorption is
important, followed by an optically thin decline.
If these are the only significant processes
and the injected power law particle spectrum does
not vary, the radio light
curves at various frequencies are self-similar.
However, as discussed in \cite{BF04}
and in \S~\ref{sp}, the action of inverse Compton cooling near optical
maximum light can depress the radio light curves in the optically
thin phase.

An additional factor noted above is that there is a minimum Lorentz
factor for the electrons, $\gamma_{min}$, and the radio emission is
affected if electrons with this energy radiate in the radio.
Combining equations (\ref{rsh}), (\ref{epe}) and (\ref{gbar}), we have
\begin{equation}
\gamma_{min}=0.3\epsilon_{e-1}E_{51}^{0.88}M_1^{-0.64}A_*^{-0.24}
t_{10}^{-0.24},
\end{equation}
which corresponds to a frequency
\begin{equation}
\nu_m=0.5\times
10^7\epsilon_{e-1}^2\epsilon_{B-1}E_{51}^{1.76}M_1^{-1.28}t_{10}^{-1.5}~{\rm Hz}.
\end{equation}
These equations are  sensible only if the parameters yield
$\gamma_{min}>1$; it can be seen that, in typcial cases, not all the
electrons are accelerated to a relativistic velocity.
The frequency $\nu_m$ is equivalent to the $\nu_m$ used in studies of GRB
afterglows.
In the supernova case, $\nu_m$ is typically  below
the observed radio wavelength region; a supernova with unusually high energy and
low mass is needed to have an observable effect of $\nu_m$.

A more important effect is the normalization of the relativistic electron
particle distribution function.
\cite{Che98} assumed that the power law distribution extends down to
$m_e c^2$ and the energy density in the relativistic component is
proportional
to the postshock energy density.
In this case, if $N(E)=N_0E^{-p}$, then $N(E)\propto t^{-2}$
and $L_{\nu}\propto R^3 N_0 B^{(p+1)/2}\propto t^{-(5+p-6m)/2}$.
In the case that $\gamma_{min}>1$, the same assumption about
the evolution of the energy density leads to
$N_0\propto R^{2(p-2)}t^{-2(p-1)}\propto t^{2(m-1)(p-1)-2m}$, and
thus $L_{\nu}\propto t^{(4mp-5p-2m+3)/2}$.
For the particular case $p=3$, which is relevant to SNe Ib/c,
we have $L_{\nu}\propto t^{5m-6}$, as opposed to $t^{3m-4}$ for the
standard case; the decline is steeper by $t^{-2(1-m)}$.

The model properties for the optically thin regime can be
compared to the observed properties in Table 1.
The spectral indices of $\alpha\approx -1.0$ correspond
to $p\approx 3$.
Over the range of ages where the observations are made, we
do not expect inverse Compton or synchrotron losses to be
important for the radio emitting electrons (\S~\ref{sp}),
implying that $p\approx 3$ is the injection spectrum.
As discussed in \S~\ref{sp}, this is steeper than the result
expected in the test particle limit ($p=2$) and may imply
that the accelerating shock wave is cosmic ray dominated.
One exception is SN 2002ap, which was observed at an early
time near the peak optical luminosity when inverse Compton
cooling may be important.
\cite{BF04} have argued that in this case
the particles are injected with $p\approx 2$ and the spectrum
is steepened to $p\approx 3$ by inverse Compton losses.
This model can explain the flatter rate of decline of
the radio emission.

The observed spectral index is in approximate accord with
expectations from the nonlinear shock theory.
\citet[][~their Fig. 8]{ER91} show that the expected
spectral index is compatible with those observed in young Galactic
supernova remnants for models in which electron acceleration
occurs to high maximum energy.
The electrons in radio supernovae radiate in a higher magnetic
field, so their energies are lower, $\gamma\sim 90$ (eq. [\ref{gam}]),
and the energy index expected in the model is $p\approx 2.8$.
\citet[][~their Fig. 2]{EBB00} approximate the electron spectrum
as power law segments with a break at $\gamma\sim 10^3$; below
the break, $p\approx 2.8$.
Thus, there is approximate accord between the nonlinear shock model
and the supernova observations.
A prediction of this model is that the observed radio spectra should
flatten with time.
However, the evolution is slow.
A factor of 10 increase in age leads to an increase in $\gamma$ by
3.2 (eq. [\ref{gam}]) and a flattening in $p$ by $\sim 0.1$
\citep[Fig. 8 of][]{ER91}.
This is not observable in currently available 
supernova radio light curves of SNe Ib/c.

The standard model for the optically thin decline, with $m=0.88$ and $p=3$,
gives $\beta=-1.3$.
An uncertainty of $\pm 0.3$ in $p$ corresponds to $\pm 0.15$
in $\beta$.
If the lowest electron energy evolves with time, the decline
steepens by 0.2.
The observed range in $\beta$ 
(Table 1) can be plausibly accounted for, although the
mechanism that is primarily responsible for the range is not
known.
One discrepant decline is the relatively flat decline of SN 2002ap,
which can be accounted for by the inverse Compton cooling
model of \cite{BF04}.

\subsection{Range of Radio Luminosity}

One of the most notable radio properties of SNe Ib/c is the large
range of radio luminosity.  
This is exemplified by the fact that SN 2003L had a higher peak
luminosity than SN 2002ap by $\sim 10^3$;  if one considered the luminosity
at the same time during the optically thin phase
($\sim100$ day), the difference would
be $\sim2\times 10^4$.
Here we investigate possible reasons for the luminosity range
of the supernovae.

We assume that the supernovae are characterized by the same value
of $m$; the possible variation in this quantity is small and we do not
expect it to contribute to the luminosity difference.
We also assume that the particle spectra are characterized by $p=3$.
If the relativistic electron spectrum extends to $m_ec^2$,
the radius of the forward shock wave
 at the time of the synchrotron self-absorption
peak is \citep{Che98}
\begin{equation}
R_p=4.0\times 10^{14}\alpha^{-1/19} \left(f\over 0.5\right)^{-1/19}
\left(F_{op}\over {\rm mJy}\right)^{9/19}  \left(D\over {\rm Mpc}\right)^{18/19}
\left(\nu\over 5 {\rm~ GHz}\right)^{-1}  {\rm~cm},
\label{rp}
\end{equation}
where $\alpha=\epsilon_e/\epsilon_B$ is the ratio of relativistic electron
energy density to magnetic energy density,
$f$ is the fraction of the spherical volume occupied by the
radio emitting region,
$F_{op}$ is the observed peak flux, and $D$ is the distance.
The results for $R_p/t_p$, where $t_p$ is the age at peak flux,
are given in Table 3,
and can be compared to expectations from the hydrodynamic model
(eq.[\ref{rsh}]).
The model shows agreement with the observations for plausible
supernova parameters; the radio emission can be attributed to the
shock region generated by the outer supernova ejecta.

Once $R_p/t_p$ is determined, the forward shock velocity can be found
from $v_s\approx 0.88 R_p/t_p$, and $\gamma_{min}$ from equations (\ref{epe})
and (\ref{gbar}).
The electron spectra extend to the nonrelativistic range,
as assumed in equation (\ref{rp}),
except for SN 2002ap, for which $\gamma_{min}=2.2$ if $\epsilon_e=0.1$. 
We have $R_p\propto \gamma_{min}^{-(p-2)/(2p+13)}\propto \gamma_{min}^{-1/19}$
($p=3$) 
\citep{Che98}, resulting in $R_p$ being reduced by 0.96 for SN 2002ap.
The effect on the velocity is  small.

The magnetic field at the time of the synchrotron self-absorption
peak is \citep{Che98}
\begin{equation}
B_p=1.1\alpha^{-4/19} \left(f\over 0.5\right)^{-4/19}
\left(F_{op}\over {\rm mJy}\right)^{-2/19}  \left(D\over {\rm Mpc}\right)^{-4/19}
\left(\nu\over 5 {\rm~ GHz}\right)  {\rm~G},
\end{equation}
where $f$ is the fraction of the spherical volume occupied by the
radio emitting region and $F_{op}$ is the observed peak flux.
This expression again assumes  that the electron energy distribution extends
down to  $\sim m_ec^2$.
For SN 2002ap, the fact that $\gamma_{min}>1$ leads to a higher value
of $B_p$ by a factor of 1.17.

Combining this $B$ field estimate with equation (\ref{ub})
leads to
\begin{equation}
A_*\epsilon_{B-1}\alpha^{8/19}=1.0 \left(f\over 0.5\right)^{-8/19}
\left(F_{op}\over {\rm mJy}\right)^{-4/19}  \left(D\over {\rm Mpc}\right)^{-8/19}
\left(\nu\over 5 {\rm~ GHz}\right)^2 t_{10}^2.
\end{equation}
The results for the observed supernovae are in Table 3.
SN 2003L stands out as having a high circumstellar density:
\begin{equation}
A_*=35\epsilon_{B-1}^{-1}\alpha^{-8/19}=35\left(\epsilon_e\over 0.1\right)^{-8/19}
\left(\epsilon_B\over 0.1\right)^{-11/19}.
\end{equation}
The maximum value for the $\epsilon$'s is $\sim 1/3$, so the implication is
that the circumstellar density around SN 2003L is high for a Wolf-Rayet star
and may require special circumstances at the end of the star's evolution.
This is consistent with the fact that SNe Ib/c are rarely as radio luminous
as SN 2003L \citep{Ber03,Sod06a}.

If we make the assumption that the difference in radio properties is due to
differences in circumstellar density ($A_*$) and that the supernovae are
similar, the velocity difference between SN 2002ap and SN 2003L can be compared
to that expected in the hydrodynamic model, $R_p/t_p\propto A_*^{-0.12}
t_p^{-0.12}$ (eq. [\ref{rsh}]).
The predicted velocity ratio is 3.2, whereas the observed is 3.3;
the high velocity in SN 2002ap can be attributed to the
low circumstellar density and early time of observation.
However, differences in the supernovae could also contribute to the differences
in velocity.
For SN 2002ap, \cite{Maz02} have estimated an
explosion energy of $(4-10)\times 10^{51}$ ergs and ejecta mass of $2.5-5\Msun$,
although these values are quite uncertain.
Estimates for SN 2003L are not available.

Overall, the range in the radio luminosities of SNe Ib/c can be
accounted for by the expected range in the circumstellar densities,
as deduced from the range of circumstellar densities in Galactic
Wolf-Rayet stars, if $\epsilon_B\sim 0.1$.
However, some contribution to the range from variations in the
$\epsilon$ efficiency parameters cannot be ruled out and may be needed
to give the broad distribution of peak luminosities of the SNe Ib/c.
As argued in \cite{Che98}, the  high magnetic fields found in SNe Ib/c
cannot plausibly be attributed to the shock compression of the
ambient stellar wind magnetic field and field amplification is
required.
One mechanism for field amplification is the turbulence generated
by hydrodynamic instabilities in the decelerating shocked shell.
However, \cite{JN96} find that $u_B$ generated in this way is $<1$\%
of the turbulent energy density, although this result may be an
underestimate due to the finite numerical resolution.
An alternative mechanism for amplification is the turbulence
generated by the diffusive shock acceleration of cosmic rays.
\cite{B04}
finds that this mechanism can generate a magnetic
energy density $u_B\sim(1/2)(v_s/c)u_{cr}$, where $u_{cr}$ is the
cosmic ray energy density.
Because we expect that $u_{cr}$ is a large part of the postshock
energy density, efficient field amplification can be accomplished
because $v_s/c\sim0.1$ for SNe Ib/c.
An implication of this scenario is that $\epsilon_B$ is not
constant, as we have assumed, because of the evolution of $v_s$.
However, the evolution is modest, $v_s\propto t^{-0.12}$.
The effect on the optically thin flux evolution is likewise modest
because $F_{\nu}\propto B^{(p+1)/2}\propto B^2$ for $p=3$.
The steepening of the flux evolution is by $t^{-0.12}$.

\section{X-RAY EMISSION}

Since the first X-ray observations of supernovae, three mechanisms
have been discussed for the emission: synchrotron radiation, thermal
emission, and inverse Compton emission \citep{CKF82,Che82b,Fra82},
and they are the mechanisms that we discuss here.
The results deduced from the radio emission provide a
starting point for the physical conditions that can be
expected.
However, the radio observations do not provide a unique model;
the combination of radio and X-ray data can be expected to better
define the physical situation in the SNe Ib/c.

\subsection{Thermal Emission}

For a given set of parameters $M$, $E$, and $A$, the hydrodynamic
interaction is well-defined, and the density and pressure profiles
are determined.
Thermal emission is dominated by emission from the reverse shock region.
If the postshock gas is in ionization and temperature equilibrium,
the temperature is
\begin{equation}
T_{eq}=5.7\times 10^8 \zeta_1 E_{51}^{0.88}\left(M\over 10\Msun\right)^{-0.64}
A_*^{-0.24}t_d^{-0.24}{\rm~K},
\end{equation}
where the gas is assumed to be fully ionized, and $\zeta_1=1/2$ for
H and $2Z/(Z+1)$ for heavy elements with charge $Z$.

The time scale for energy transfer from ions to electrons is
\begin{equation}
t_{e-i} = {4.2\times 10^{-22} \over \ln \Lambda} \zeta_2(Z) 
{T_e^{3/2} \over \rho},
\end{equation}
where $\zeta_2 = 1$ for H and $\zeta_2 = 4/Z$ for heavier elements of
charge $Z$ and $\ln \Lambda \approx 30$ is the Coulomb logarithm. For
the reverse shock this can be written as
\begin{equation}
t_{e-i} = 51 \zeta_2(Z) \left({v_s \over 5\times 10^4 \kms}\right)^{2}
    \left({T_e \over 10^8 ~ {\rm K}}\right)^{3/2} A_*^{-1} t_{10}^2 ~~ \rm days.
\end{equation}
The reverse shock temperature is $\ga 5 \times 10^8$ K and it is
clear that the electrons and ions can not be kept in equipartition by
Coulomb collisions only. We can reverse this and estimate the
minimum electron temperature:
\begin{equation}
T_{e} = 3.2 \times 10^7 \zeta_2(Z)^{-2/3} \left({v_s \over 5\times 10^4
\kms}\right)^{-4/3} 
A_*^{2/3} t_{10}^{-2/3} ~~ \rm K.
\end{equation}
For an oxygen dominated ejecta, the typical temperature is 
$(0.5 - 2)\times 10^8$ K; a helium dominated composition will give a
temperature lower by a factor $\sim 2.5$. It is plausible that plasma
instabilities in the shock lead to a higher value, as is the case
for supernova remnants. Complete equipartition is, however, unlikely. 

We have also estimated the timescale for collisional ionization
behind the reverse shock,
which can be written 
\begin{equation}
t_{ion} = {1.2\times 10^{-11} \over C_i(T_e)} \left({v_s \over 5\times 10^4
     \kms}\right)^{2} A_*^{-1} t_{10}^2 ~~  \rm days,
\end{equation}
where $C_i(T_e)$ is the collisional ionization rate for ion $i$, and
we have assumed a composition dominated by helium or heavier
elements. Assuming a temperature given by the pure Coulomb heating
case we find that, although the timescale is strongly dependent on
the composition, intermediate mass elements like oxygen will in
general be completely ionized, while heavy elements like iron will be
in their He-like stages, i.e. under-ionized with respect to the
temperature.
For the Type Ic SN 1994I, \cite{Iwa94} propose an
outer layer with a composition X(O)=0.43, X(C)=0.45, and X(He)=0.09.
On the other hand, SNe Ib have He lines in their optical spectra,
and \cite{Bra02} find that there is typically evidence
for H in the highest velocity layers.
The expected composition should lead to complete ionization.

If the gas is completely ionized, the free-free emission
from the reverse shocked gas can be estimated \citep{CF03}
\begin{equation}
L_{ff}=3\times 10^{35}{(n-3)(n-4)^2\over 4(n-2)}\beta^{1/2}\zeta_2^{-1} A_*^2 t_{10}^{-1}{\ergs}
=2.4\times 10^{36}\beta^{1/2}\zeta_2^{-1} A_*^2 t_{10}^{-1}\ergs,
\end{equation}
where $\beta=T_e/T_{eq}$ 
and the second expression is for $n=10$.
Over the timescales of interest for the X-ray observations,
circumstellar X-ray  absorption is expected to be negligible for SNe Ib/c.

Substitution of the values of $A_*$ from Table 3 shows that
the predicted thermal X-ray luminosity falls below the observed
luminosity by orders of magnitude in every case.
\cite{Imm02} discussed a thermal interpretation of the X-ray
emission from SN 1994I, but they required $\dot M=10^{-5}\ml$
for an assumed $v_w=10\kms$, or $A_*=100$.
This can be compared to $\epsilon_{B-1}A_*\approx 2.7$ deduced here.
Although a low value of $\epsilon_{B-1}$ could bring up $A_*$
to the required value, the mass loss rate would be considerably
higher than that expected for a Wolf-Rayet star.
In view of this, we explore nonthermal mechanisms for the
X-ray emission.

\subsection{Inverse Compton X-Ray Emission}

One type of inverse Compton emission is  scattering of photospheric
photons with thermal hot electrons in the shock interaction region
\citep{Fra82}.
\cite{Sut03} suggested that this mechanism is responsible
for the X-ray emission observed from SN 2002ap.
However, the required electron optical depth and position of the
hot electrons are not compatible with the hydrodynamic situation
expected in SN 2002ap, and  inverse Compton scattering of photospheric
photons by relativistic electrons 
is more plausible \citep{BF04}.
The observed radio emission from a supernova gives information
on the spectrum of relativistic electrons.

The radio observations of SNe Ib/c show that the power law electron
spectrum is well approximated by $p=3$.
For this case, the inverse Compton X-ray luminosity is
\citep{CFN06}
\begin{equation} 
\frac{dL_{\mathrm{IC}}}{dE } 
\approx 8.8\times 10^{36}  \epsilon_{r-1} \gamma_{\rm min} E_{\rm
keV}^{-1} ~A_* 
v_{s4} \left(L_{\rm bol}(t) \over 10^{42} \ergs
\right)
t_{10}^{-1}~{\rm~ ergs~s^{-1}~keV^{-1}},
\label{eq11}
\end{equation}
where $\gamma_{\rm min}$ is the minimum Lorentz factor of the
relativistic electrons, $v_{s4}$ is the forward shock velocity in
units of $10^4\kms$, and $L_{\rm bol}$ is the bolometric luminosity
of the supernova.  
Instead of $A_*$, the radio emission gives the quantity
$S_*\equiv \epsilon_{B-1}A_* \alpha^{8/19}$, so that equation (\ref{eq11})
becomes
\begin{equation} 
E\frac{dL_{\mathrm{IC}}}{dE } 
\approx 8.8\times 10^{36}   \gamma_{\rm min}  ~S_* \alpha^{11/19}
v_{s4} \left(L_{\rm bol}(t) \over 10^{42} \ergs
\right)
t_{10}^{-1}~{\rm~ ergs~s^{-1}}.
\label{eq12}
\end{equation}
The epsilon parameters now enter only through the ratio $\alpha=\epsilon_e/\epsilon_B$.

Predictions of the X-ray luminosity using equation (\ref{eq12}) are
given in Table 4, where the values of $v_{s}$ have been taken from the
velocity at peak radio luminosity (Table 3) and evolved to the time
of the X-ray observation.
The value of $\alpha$ is assumed to be 1.
The values of $L_{bol}$ for SN 2002ap, 2003L, and 2003bg are given in
the references listed in Table 3.
The early values for SN 1994I are from \citet{Ric96}.
No direct optical observations of SN 1994I at an age of 7 years
are available, but estimates of
the luminosity are possible by analogy to SN 1987A.
At an age of 7 years, SN 1987A had faded by 15 magnitudes from maximum light
\citep{Sun03}, to $\sim 2\times 10^{36}\ergs$.
The power input is presumably from $^{44}$Ti at this age and SNe Ib/c are
estimated to have $\la 10^{-4}\Msun$ of $^{44}$Ti synthesized in the
explosion \citep{Tim96}, comparable to the $\sim10^{-4}\Msun$ of $^{44}$Ti
estimated for SN 1987A.
In addition, the estimated mass of $^{56}$Ni, $0.07\Msun$ \citep{YBB95},
is comparable to that ejected in SN 1987A.
The late luminosity of SN 1994I, is probably considerably less that
$2\times 10^{36}\ergs$ because the ejected mass in SN 1994I, $\sim1\Msun$ \citep{YBB95},
is much less than that of SN 1987A and the ejecta are less able to
absorb the radioactive power.
There is not a published optical light curve for SN 2001ig, so we used
amateur photometric measures \citep{Pel04} to estimate
the bolometric luminosity for this case.

The results show varying degrees of agreement with the observed X-ray luminosities.
The observation that is most consistent with the expectations from inverse
Compton is the early observation of SN 2002ap, which is in agreement with
the interpretation of \cite{BF04}.
The other predictions of inverse Compton emission are lower than expected,
and require high values of $\alpha=\epsilon_e/\epsilon_B$ to come
into agreement with the observed luminosities.
A problem with departures from energy equipartition of magnetic fields and
electrons ($\alpha\ne 1$) is that we argued for $\epsilon_e\approx \epsilon_B
\approx 0.1$ in order to approximately reproduce the range of Wolf-Rayet
wind densities.
Because the $\epsilon$'s are expected to be $\la 1/3$, the possible
departure from equipartition is limited; this argument especially applies
to the radio luminous supernovae (SN 2003L and SN 2003bg).

A general expectation of the inverse Compton model is the relation between
the optical luminosity and the inverse Compton luminosity (eq. [\ref{eq12}]);
this relation is not shared by other mechanisms (thermal and synchrotron).
The existing X-ray observations do not provide good light curve coverage,
but, where multiple epochs do exist, they do not show the expected relation.
The large decline expected between days 30 and 120 for SN 2003bg does not
appear in the observed fluxes, which argues against the inverse Compton
mechanism, at least at the late time.
For SN 1994I, the inverse Compton definitely fails to explain the late
time observations, as will be shown explicitly in \S \ref{sec_synch}.
Searches for inverse Compton emission should concentrate on observations
around the time of maximum optical light.

\subsection{Synchrotron}
\label{sec_synch}
Because of the lack of a detailed theory for the acceleration
of relativistic electrons, the X-ray synchrotron emission is
typically estimated by extending the radio synchrotron emission
into the X-ray regime.
In view of the relatively steep spectra of SNe Ib/c and further
steepening expected from synchrotron losses, a simple extrapolation
of the optically thin radio emission leads to a low X-ray luminosity.
Table 2 gives the value of $\nu L_{\nu}$ for the various supernovae
at 5 GHz; in cases where there are no radio observations at the time
of the X-ray observations or the emission is optically
thick at 5 GHz, a rough extrapolation of the radio
evolution has been made.
With $L_{\nu}\propto \nu^{-1}$, as is approximately observed in
these supernovae, the quantity $\nu L_{\nu}$ stays
constant with frequency.
It can be seen that the X-ray luminosity is above this extrapolation (Table 4);
if synchrotron losses to the electron spectrum were included, the
discrepancy would be larger.
It is for this reason that the synchrotron mechanism is generally not
found to be viable for X-ray supernovae \citep[e.g.,][on SN 2003L]{Sod05}.

However, as described in \S~\ref{sp}, the nonlinear theory of particle acceleration
predicts that the electron spectrum becomes flatter for $\gamma\ga 10^3$.
We use an approximate version of this theory to estimate whether
synchrotron emission can account for the late emission from
SN 1994I.
Following \citet{EBB00}, we assume power law segments for the electron energy
distribution.
Below $\gamma=10^3$, we take $p=3$, and for $\gamma> 10^3$, $p=2$.
This injection spectrum for the particles extends to X-ray emitting
particles, as argued above.
The frequency corresponding to the spectral break is
$\nu_{br}=3.4\times 10^{12}B$ Hz.
For the situation here, we have $B\propto t^{-1}$, which also
gives the time dependence of $\nu_{br}$.
For the magnetic field estimated in SN 1994I (Table 3),
$\nu_{br}=6.5\times 10^{12}t_{10}^{-1}$ Hz.

The injection spectrum is modified by inverse Compton and synchrotron
losses;
at the late times of interest here, synchrotron losses are expected
to be the dominant mechanism.
The frequency at which synchrotron losses become important, $\nu_c$,
can be found from equation (\ref{ts}).
For the magnetic field estimated for SN 1994I, we have
$\nu_{c}=4.1\times 10^{11}t_{10}$ Hz.
For these conditions, $\nu_c>\nu_{br}$ provided that $t>40$ days.
At the times of interest, the particle spectrum steepens to
$p=3$ for $\nu>\nu_c$.
We then have
\begin{equation}
{\nu L_{\nu,X-ray}\over \nu L_{\nu,radio}}=
\left(\nu_c\over\nu_{br}\right)^{1/2}\propto t.
\label{xray}
\end{equation}
For the parameters relevant to SN 1994I at an age of 2500 days,
$\nu L_{\nu,X-ray}/ \nu L_{\nu,radio}=62$.
The synchrotron mechanism is thus able to approximately account
for the late X-ray emission from SN 1994I in the context of a
particle spectrum formed in a cosmic ray dominated shock.
Equation (\ref{xray}) shows that, with the synchrotron mechanism, the evolution
of $L_{X-ray}$ is flatter than $L_{radio}$ by one power of $t$.
For the radio evolution observed for SN 1994I (Table 1), the
predicted evolution is $L_{X-ray}\propto t^{-0.3}$.

 Although this gives an estimate of the expected emission, we have
carried out more detailed calculations of the effects of synchrotron
and inverse Compton radiation on the particle spectrum and emission,
using the methods of \cite{FB98}.
 One calculation was to assume a
particle injection spectrum with $p=3.1$, normalized to reproduce
the radio emission from SN 1994I, and a bolometric supernova optical
light curve like that of SN 1994I, taken from \citet{Ric96}.
 The
optical emission is important for inverse Compton effects.
 We assume
$\epsilon_e = \epsilon_B = 0.1$, $A_*=3$, $v_s(t)=6 \times 10^4 (t/10
\rm \ days)^{-1/8} \kms$ (i.e., $n=10$).  This model fails to
reproduce the late X-ray emission from SN 1994I 
 by a large factor
(Fig. 1).
 In another model, we include an injection spectrum
appropriate to
 a cosmic ray dominated shock wave, as calculated by
\citet[][~ their Fig. 1]{EBB00}.
 The flattening of the spectrum to
high energy greatly enhances
 the X-ray synchrotron emission
(Fig. 1), while the effect on
 the radio emission is minor.
Finally, we show this same calculation, but with a supernova
 light
curve appropriate to SN 1997ef, which had a broader optical maximum
than SN 1994I \citep{Iwa00}.
 The effect on the inverse Compton X-ray
emission near maximum light
 can be seen (Fig. 1), and it is clear
that inverse Compton emission is important only during the first $\sim
50$ days even for a boad optical light curve such as SN 1997ef.
These results confirm that the cosmic ray dominated shock model
 can
explain the observed X-ray emission on day $\sim2500$.
 The model is
also consistent with the observed upper limit on day
 52, but falls
below the detection on day 82.
 However, the early detection of SN
1994I did not have high
 statistical significance: $3.4\sigma$ in a
6.4 ks observation
 and $3.1\sigma$ in a 29.9 ks observation with
ROSAT \citep{Imm98}.
 
 Fig. 1 shows that the X-ray light curves
have an imprint of the
 optical light curve due to inverse Compton
radiation near
 maximum light.
 Selected spectra show the basic
components of the multiwavelength spectrum: optical photospheric
emission, inverse Compton emission, and synchrotron radiation
(Fig. 2). At 10 days there is a strong 
inverse Compton component extending from the ultraviolet
to the X-rays. At 30 days this has almost disappeared. The curvature
of the synchrotron component is apparent, especially at 10 days when
 the radio range is also affected by cooling.  
 For SN 1994I, we
cannot rule out the possibility of late dense circumstellar
interaction giving rise to thermal X-ray emission because there
 are
not simultaneous observations at other wavelengths.
 In our model,
this would require dense matter at some distance
 from the supernova,
as appears to be the case in SN 2001em
 \citep{CC06}.  Our
suggestion of late synchrotron emission from a Type Ic like SN 1994I
needs confirmation from more complete X-ray observations
 of a
similar event. Spectral observations would provide the most direct
constraints, but as Fig. 1 shows,  the evolution of the X-ray
light curve can also discrimnate between different mechanisms.  
 
\subsection{SN 1998bw}

None of the supernovae discussed so far have a clear connection with
GRBs and the emission can be accounted for by an interaction region
driven by the fast outer supernova ejecta.
SN 1998bw was observed as a bright SN Ic and was associated with
GRB 980425.
The early turn-on of the radio emission suggests mildly relativistic
expansion in the synchrotron self-absorption model \citep{Kul98,LC99};
this implies that the nonrelativistic analysis presented here may
not be accurate, but it can still be expected to give approximate
results.
The early peak emission (42 mJy 5 GHz flux at 13 days) implies
$\epsilon_{B-1}A_*\approx 0.1$.
For a comparable efficiency of magnetic field production, the circumstellar
density is low for SN 1998bw compared to the typical SN Ib/c.

SN 1998bw has been observed as an X-ray source over days $1-1200$, showing
a remarkably low rate of decline \citep{Kou04} (a factor $\sim10$ over
this time period).
Inverse Compton scattering of photospheric photons would decline much
more rapidly after maximum light, so synchrotron emission is the most
likely emission mechanism.
The radio emission has an optically thin spectral index $\nu^{-0.75}$,
implying $p\approx 2.5$.
If the observed flux at $t=100$ days is extended to the X-ray range, the
luminosity falls short of the observed X-ray luminosity by a factor of $\sim10$.
Considering that synchrotron losses can further decrease the X-ray
flux by 20, the radio through X-ray emission is not consistent with
injection of electrons with a single power law index.
As discussed above, there are reasons to believe that the energy
index might flatten at high energy if the shock waves are cosmic ray
dominated.
In this case, the synchrotron X-ray emission can decline more slowly
than the radio emission (Fig. 1), as is observed in SN 1998bw.
Alternatively, the X-ray  emission could originate from a different 
region from the radio.
\citet{Kou04} discuss the possibility that the X-ray emission is
related to off-axis emission from a fast jet related to the gamma-ray burst.

\section{DISCUSSION AND CONCLUSIONS}

We have found that the radio and X-ray emission from SNe Ib/c can
be accounted for by supernova interaction with the wind from the
Wolf-Rayet star progenitor.
Following the arguments of \cite{Che98},
synchrotron self-absorption is generally responsible for the 
early absorption of the radio emission.
The radio synchrotron emission from most of the SNe Ib/c considered 
here imply an electron energy index $p\approx 3$.
The radio spectral index of SN 1998bw is distinctly flatter, with
$p\approx 2.5$.
The only other SN Ib/c with a relatively flat spectrum is SN 2002ap,
with $p\approx 2$ in the model of \citet{BF04}.
Interestingly, SN 2002ap has the highest velocity of any of the supernova
in our group of normal Ib/c events.
It may be that as the shock velocity increases and relativistic effects
become important, there are changes in the character of the particle acceleration.

The SNe Ib/c provide the opportunity to examine particle acceleration at
shock velocities ($\sim0.1c$) intermediate between those present in Galactic supernova
remnants and those in GRBs.
Although the properties of the synchrotron radio emitting particles are
clear ($p\approx 3$ is typical), the properties of the X-ray synchrotron
emission are uncertain.
If the circumstellar densities are typical of Wolf-Rayet stars,
thermal radiation cannot explain the observed X-ray emission and
a nonthermal mechanism is required.
Although inverse Compton 
scattering of photospheric photons might be able to explain X-ray emission
close to maximum optical light, it generally fails to explain late emission.
If this emission is synchrotron radiation, there must be a flattening of
the particle spectrum at high energy, as can occur in diffusive acceleration
in cosmic ray dominated shock waves.
More detailed X-ray observations are needed to confirm this picture.
We expect the early X-ray luminosity to be related to the optical luminosity
of the supernova if it is inverse Compton emission; 
as the supernova fades, synchrotron emission may become
the dominant mechanism, with a relatively slow rate of decline.

An additional piece of information from our interpretation is the magnetic field
in the shocked region.  If the circumstellar densities are typical of Wolf-Rayet
stars, we find that efficient amplification of magnetic fields is needed,
with $\epsilon_B\sim0.1$.
A possible reason for the field is a streaming instability that accompanies
cosmic ray acceleration \citep{B04}, which may explain the high amplification
at the high shock velocities present in SNe Ib/c.
With this mechanism, $\epsilon_B\propto v_s$, so the value of $\epsilon_B$ would
be smaller in Galactic supernova remnants, as is observed.

\acknowledgments
This research was partially carried out while the authors were visiting the
Kavli Institute of Theoretical Physics, supported in part by the National
Science Foundation under Grant No. PHY99-07949.
In addition, this research was supported in part by Chandra grant TM4-5003X
and NSF grant AST-0307366 and the Swedish Research Council and Swedish
National Space Board.

\clearpage

\begin{center}
\noindent{{Table 1.} Optically Thin Radio Properties of Type Ib/c Supernovae}

\begin{tabular}{ccccc}
\hline
Supernova & $\alpha$ & $\beta$  &  Age &   References \\
        &        &         &    (days)  &     \\
\hline
1983N     &    $-1.0$ & $-1.6$   &  $30-300$   &  1 \\
1984L     &    $-1.0$ & $-1.5$   &  $100-200$   &  2  \\  
1990B     &    $-1.1$ & $-1.3$   &  $70-200$   &  3  \\
1994I     &    $-1.0$ &  $-1.3$  & $20-800$    &  4  \\
2001ig     &   $ -1.06$ & $-1.5$   &  $70-700$   &  5  \\
2002ap &   $-0.9$ &  $-0.9$  &  $4-20$   &  6  \\
2003L &   $-1.1$  &  $-1.2$  &  $100-400$     &  7  \\
2003bg &   $-1.1$  &  $-2$  & $60-1000$      &  8  \\
\hline
\end{tabular}
\end{center}
\noindent{The references to the data are:
(1) Weiler et al. 1986; (2)  Panagia et al. 1986;
(3) Van Dyk et al. 1993;
(4)  Stockdale et al. 2005; (5) Ryder et al. 2004;
(6) Berger et al. 2002; (7) Soderberg et al. 2005;
(8) Soderberg et al. 2006b}

\clearpage

\begin{center}
\noindent{{Table 2.} X-ray Observations of Type Ic Supernovae}

\begin{tabular}{cccccc}
\hline
Supernova & Distance & Age  &  Luminosity &  $\nu L_{\nu}$ (5 GHz)  &  References \\
   & (Mpc)  &(days)& (10$^{38}$~ergs s$^{-1}$)& (10$^{38}$~ergs s$^{-1}$)  &     \\
\hline
1994I &    8.3 & 52,82  &  $<1, 1.7$ (0.3--2 keV)   & 0.04 &  1,2  \\
"    &     &   2271,2637  &  0.17,0.14 (0.3--2 keV)   & 0.001 &  2  \\
2001ig &   12.3 &  170,191  & 1,0.6 (0.2--10 keV)  & 0.1 &  3,4   \\
2002ap &   10.4 &  4  & 0.9 (0.3--10 keV)    & 0.003 &  5,6  \\
2003L &   96 &  40  &   90 (2--10 keV)    & 13  &  7,8  \\
2003bg &   19.5 &  30,120   &   44,9  (0.3--10 keV)    & 7,3  &  9,10   \\
\hline
\end{tabular}
\end{center}
\noindent{The references to the data are:
(1) Immler et al. 1998; (2)  Immler et al. 2002;
(3) Schlegel \& Ryder 2002;  (4) Ryder et al. 2004;
(5)  Rodriguez Pascual et al. 2002; (6) Sutaria et al. 2003;
(7) Kulkarni \& Fox 2003;
(8) Soderberg et al. 2005;
(9) Pooley \& Lewin 2003;
(10) Soderberg et al. 2006b}
\vspace{0.1 in}

\clearpage

\begin{center}
\noindent{{Table 3.} Properties of Type Ib/c Supernovae in a Synchrotron Self-Absorption Model}

\begin{tabular}{cccccccc}
\hline
Supernova &    $t_p$ & $\nu_p$ &  $F_{op}$ & $D$  & $B_p$ & $\epsilon_{B-1}A_*\alpha^{8/19}$ & $R_p/t_p$  \\
        &  (days)   & (GHz) &  (mJy)  &  (Mpc)&  G  &   &   (km s$^{-1}$) \\
\hline
1983N    &    21   &  4.88  &  18 &  5.1  &  0.56 &   1.15  & 42000  \\
1990B     &    91   & 1.49  &  1.6  &  18.0  & 0.17 &   2.0  & 33000  \\
1994I     &    36   &  4.86  & 17  &  8.3  &  0.51 &   2.8   &  38000 \\
2001ig     &   42  & 4.8    &  18  &  12.3   & 0.46  &  3.1  & 49000  \\
2002ap &   8      &  1.43  &  0.3 &  10.4 & 0.26 &  0.04   & 105000 \\
2003L &   170    &  4.9   &  2.4  &   96  & 0.38  &   34    &  32000 \\
2003bg &  60   &  8.46  &  40  &   19.5  &  0.68 &   13    &  44000 \\
\hline
\end{tabular}
\end{center}

\clearpage

\begin{center}
\noindent{{Table 4.} Expectations for Inverse Compton X-ray Luminosity}

\begin{tabular}{cccccccc}
\hline
Supernova &  Age  &  $L_{bol}$ & $v_s$ & $\gamma_{min}$ &  $EdL_{IC}/dE$  &  $L_{obs}$  \\
   &     (days)& (10$^{42}$~ergs s$^{-1}$)&  ($10^4\kms$)&  & (10$^{38}$~ergs s$^{-1}$)  & (10$^{38}$~ergs s$^{-1}$)     \\
\hline
1994I &    52  &  0.2  &   3.2 & 1 &  0.03 &  $<1$  \\
1994I &    82  &  0.1   &  3.0  & 1  & 0.009  &  1.7   \\
1994I &     2271 &  2(-6)   &   2.0  & 1 &  4(-9) &  0.17  \\
1994I &     2637  &  2(-6)   &  2.0  & 1 &  4(-9) &  0.14  \\
2001ig &    170  &  0.15  &    3.6  & 1 & 0.009 &  1   \\
2001ig &    191  &  0.1  &  3.6  & 1   & 0.005 &  0.6   \\
2002ap &    4  & 1.5    & 10.0  & 2.5  & 0.2 &  0.9 \\
2003L &     40  &   1.6  &   3.3   & 1 &  4  &  90  \\
2003bg &  30    &  2.3    &   4.2 & 1 & 4  &   44  \\
2003bg &  120    &  0.3    &   3.6  & 1 &  0.1   &   9  \\
\hline
\end{tabular}
\end{center}

\clearpage

\begin{figure}[!hbtp]
\epsscale{.80}
\plotone{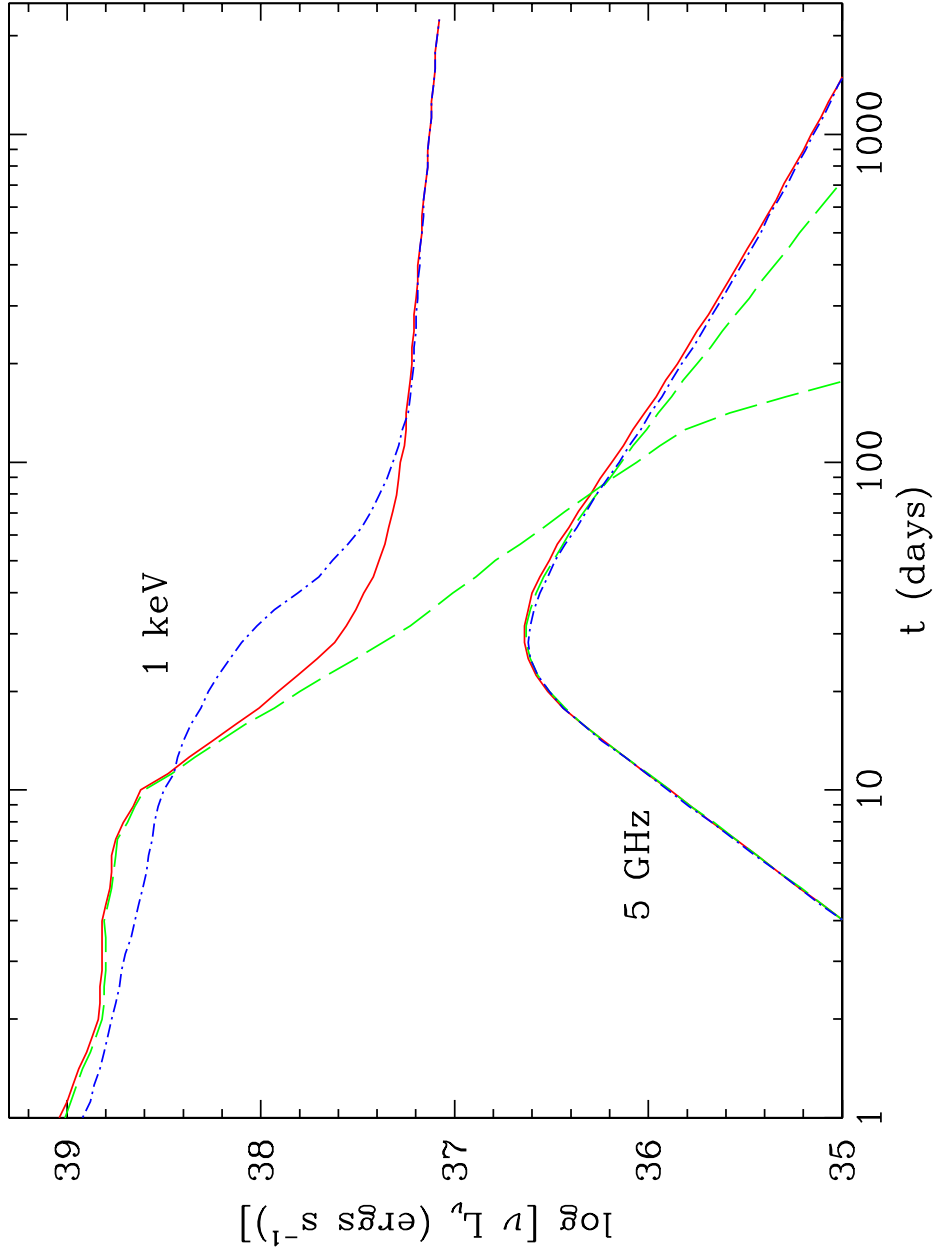}
\caption{Model radio and X-ray light curves, including synchrotron
and inverse Compton radiation.
The green curve ({\it dashed line}) has a power law injection particle spectrum with
$p=3.1$ and assumes an optical light curve like that of SN 1994I.
The red curve ({\it solid line}) has an injection spectrum based on acceleration in
a cosmic ray dominated shock and assumes a supernova like SN 1994I.
The blue curve ({\it dashed-dot line}) is the same as the red, but assumes a supernova like
SN 1997ef.
\label{lt_curve}}
\end{figure}

\begin{figure}[!hbtp]
\epsscale{.80}
\plotone{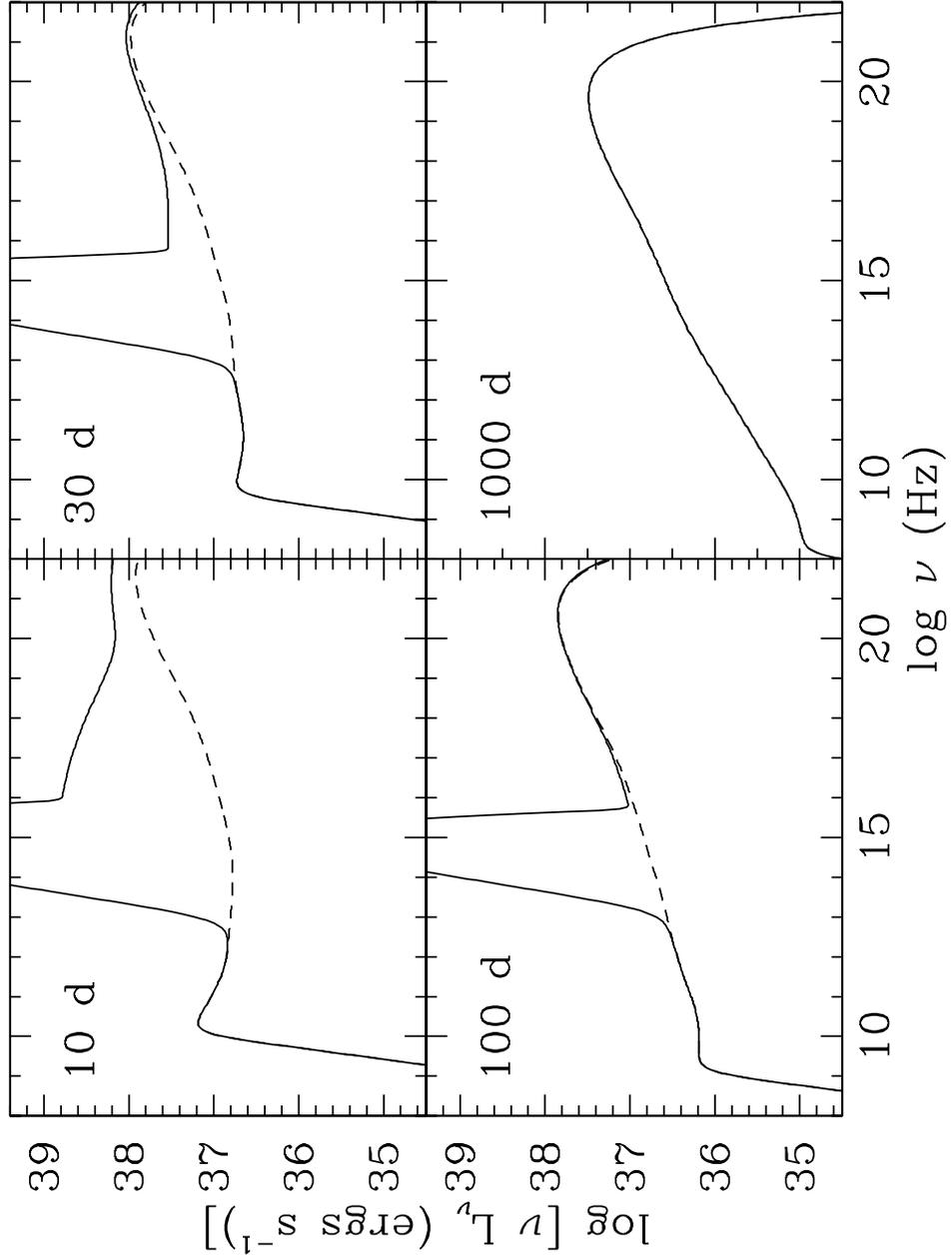}
\caption{Selected spectra for an injection spectrum based on acceleration in
a cosmic ray dominated shock and  a supernova like SN 1994I. The
dashed line shows the synchrotron component.
}
\end{figure}

\end{document}